\documentclass[preprint,amsmath,amssymb,aps]{revtex4}
\usepackage{graphicx}
\usepackage{bm}
\usepackage{subfigure}
\usepackage{amsmath}
\usepackage[utf8]{inputenc}

\newcommand{\pr}{Phys. Rev.\ }

\newcommand{\jpa}{J. Phys. A\ }

\newcommand{\etals}{{\em et al.}}  

\begin{document}

\title{Quantum difference parametric amplification and oscillation}

\author{M.~K. Olsen}
\affiliation{School of Mathematics and Physics, University of Queensland, Brisbane, 
Queensland 4072, Australia.}

\date{\today}

\begin{abstract}

A recent article [W.C.W. Huang and H. Batelaan, arXiv:1708.0057v1] analysed the dualism between optical and difference parametric amplification, performing a classical analysis of a system where two electromagnetic fields are produced by another of a frequency which is the difference of the frequency of the other two. The authors claimed that this process would not violate energy conservation at the classical level, but that a quantum description would necessarily require a non-Hermitian Hamiltonian and therefore would not exist. In this work we show that the process can proceed quantum mechanically if described by the correct Hamiltonian, that energy conservation is not violated, and that fields are produced with interesting quantum statistics. Furthermore, we show that the process can be thought of as different types of already known three-wave mixing processes, with the actual type depending on either initial conditions or personal preference.  

\end{abstract}

\maketitle

\section{Introduction}
\label{sec:intro}

Optical parametric processes wherein modes of the elctromagnetic field interact via a $\chi^{(2)}$ nonlinear medium have a long history in quantum physics, with the theory having first been developed by Armstrong \etals~\cite{Armstrong}. These processes, which can also be thought of as three-wave mixing, are extremely versatile. The most commonly known and useful for quantum information purposes is possibly the non-degenerate parametric oscillator, used by Ou \etals~\cite{Ou} to demonstrate the Einstein-Podolsky-Rosen (EPR) paradox~\cite{EPR}, via the Reid EPR correlations~\cite{EPRMDR}. Other well known processes are second harmonic generation~\cite{SHG}, degenerate optical parametric oscillation~\cite{OPO}, and sum frequency generation~\cite{sumfreq}. The achievement of Bose-Einstein condensation also saw such processes proposed with atoms, for both molecular association~\cite{PDDsuper,JJsuper,BECstates} and dissociation~\cite{KVKdiss}. Just as with the optical processes, these atomic parametric processes exhibit interesting quantum properties.

In a recent article~\cite{difference}, there was a proposal for a new type of parametric $\chi^{(2)}$ process, which the authors named difference parametric amplification. This process involves three electromagnetic fields, at frequencies $\omega_{1}$, $\omega_{2}$, and $\omega_{d}$, where $\omega_{d} = \omega_{2}-\omega_{1}$. After a classical analysis, the authors wrote a quantum Hamiltonian and made the claim that the process had no quantum mechanical equivalent because the Hamiltonian was non-Hermitian.

There are in fact several ways of looking at this system. It can be thought of as a Jaynes-Cummings model~\cite{JC} for light, in which the fields at the upper and lower frequencies take the place of two atomic levels coupled by the field at the difference frequency. It can be thought of as asymmetric sum frequency generation~\cite{SFG} when the lower and difference fields are pumped. Finally, when the upper and difference fields are pumped, it is equivalent to non-degenerate downconversion with an injected signal~\cite{injectOPO}.   

In this article we show that a Hermitian Hamiltonian can be written which describes the system, and that the process  will not proceed with only the field at $\omega_{d}$ initially populated. Energy conservation does not allow this. We show how the interaction can proceed as long as one of the other modes has at least some initial population, and perform a fully quantum analysis of the system in both the travelling wave and intracavity configurations, calculating both the mean fields and bipartite quantum correlations for the detection of inseparability and entanglement.

\section{Interaction Hamiltonian and Dynamics}
\label{sec:Ham}

The physical system consists of three electromagnetic modes at frequencies $\omega_{1}$, $\omega_{2}$ and $\omega_{d}$, where $\omega_{d}=\omega_{2}-\omega_{1}$. In the classical treatment by Huang and Batelaan~\cite{difference}, the mode at $\omega_{d}$ creates modes at the other two frequencies. In a quantum treatment, this obviously violates energy conservation, and when they write their quantum Hamiltonian (Eq. 18), they find it is non-Hermitian. In fact, they find that their Hamiltonian is not Hermitian under the conditions we will work with in this paper. However, the Hamiltonian they propose is different from that normally used for parametric three-wave mixing processes in a quantum analysis. 

We can write a Hermitian interaction Hamiltonian for the same physical process which is identical to those of the well known processes of non-degenerate downconversion and sum frequency generation. The physical interaction links the two fields at $\omega_{1}$ and $\omega_{2}$ via a nonlinear $\chi^{(2)}$ material and the field at $\omega_{d}$. It is instructive to consider the Jaynes-Cummings model, where two states of a two-level atom are coupled by a single mode of the electromagnetic field. When the atom is not present in one of these states, the Hamiltonian describes exactly nothing physically. This is the case we find here. If there are no photons initially present at either $\omega_{1}$ or $\omega_{2}$, the light at $\omega_{d}$ passes straight through the nonlinear material and couples two vacuum states. This is not particularly exciting, nor worthy of analysis.  

We begin with the interaction Hamiltonian in a rotating frame, which can be written as
\begin{equation}
{\cal H}_{int} = i\hbar\kappa\left( \hat{a}_{1}^{\dag}\hat{a}_{d}^{\dag}\hat{a}_{2}-\hat{a}_{1}\hat{a}_{d}\hat{a}_{2}^{\dag}\right),
\label{eq:intham}
\end{equation}
where $\hat{a}_{j}$ is the bosonic annihilation operator for the mode at $\omega_{j}$ and $\kappa$ represents the effective $\chi^{(2)}$ nonlinearity. We clearly see that this Hamiltonian is Hermitian and that similar Hamiltonians have been used to describe many parametric processes in the past. In fact, a similar Hamiltonian can be used to describe spontaneous emission from bosonic atoms~\cite{bosespon}, when only the high frequency mode is initially populated.

We can immediately see that our Hamiltonian is Hermitian, i.e. ${\cal H}_{int}^{\dag}={\cal H}_{int}$, whereas the quantum Hamiltonian given by Huang and Batelaan leads to ${\cal H}_{int}^{\dag}=-{\cal H}_{int}$. There is only one possible reason for this to happen, and that is that the energy described by their Hamiltonian is identically zero and will not lead to any interesting dynamics of the fields. As we will show below, this is not the case for our choice of interaction Hamiltonian.

Following the usual procedures~\cite{DFW}, we map the Hamiltonian, via von Neumann and Fokker-Planck equations, onto a set of It\^o calculus~\cite{SMCrispin} stochastic differential equations in the positive-P representation~\cite{P+},
\begin{eqnarray}
\frac{d\alpha_{1}}{dt} &=& \kappa\alpha_{d}^{+}\alpha_{2}+\sqrt{\frac{\kappa\alpha_{2}}{2}}(\eta_{1}+i\eta_{3}), \nonumber \\
\frac{d\alpha_{1}^{+}}{dt} &=& \kappa\alpha_{d}\alpha_{2}^{+}+\sqrt{\frac{\kappa\alpha_{2}^{+}}{2}}(\eta_{2}+i\eta_{4}), \nonumber \\
\frac{d\alpha_{d}}{dt} &=& \kappa\alpha_{1}^{+}\alpha_{2}+\sqrt{\frac{\kappa\alpha_{2}}{2}}(\eta_{1}-i\eta_{3}), \nonumber \\
\frac{d\alpha_{d}^{+}}{dt} &=& \kappa\alpha_{1}\alpha_{2}^{+}+\sqrt{\frac{\kappa\alpha_{2}^{+}}{2}}(\eta_{2}-i\eta_{4}), \nonumber \\
\frac{d\alpha_{2}}{dt} &=& -\kappa\alpha_{1}\alpha_{d}, \nonumber \\
\frac{d\alpha_{2}^{+}}{dt} &=& -\kappa\alpha_{1}^{+}\alpha_{d}^{+}.
\label{eq:Pplustravel}
\end{eqnarray}
In the above, stochastic averages of the variables $\alpha_{j}$ represent normally ordered operator expectation values in the sense that $\overline{\alpha_{j}^{+\, m}\alpha_{k}^{n}}\rightarrow\langle\hat{a}_{j}^{\dag\,m}\hat{a}_{k}^{n}\rangle$, and the $\eta_{j}$ are Gaussian noise terms with the properties $\overline{\eta_{i}}=0$ and $\overline{\eta_{j}(t)\eta_{k}(t')}=\delta_{jk}\delta(t-t')$. These equations can be solved numerically, or in a linearised form around classical mean-field solutions. However, since the latter process has previously been found to be unreliable in other parametric processes~\cite{revive,QND,SFG}, we will not pursue it here.

We can immediately see that the above equations, with only $\alpha_{d}(0)\neq 0$, will give $\alpha_{1}=\alpha_{2}=0\,\forall t$. By way of comparison, this would be a Jaynes-Cumming model with no atoms, or free propagation of the field at $\omega_{d}$. However, when we also let one of $\alpha_{i}(0)$ or $\alpha_{2}(0)$ equal to a finite value, we see that $\alpha_{d}$ couples these two fields and causes a periodic population transfer. 

\begin{figure}[tbhp]
\includegraphics[width=0.75\columnwidth]{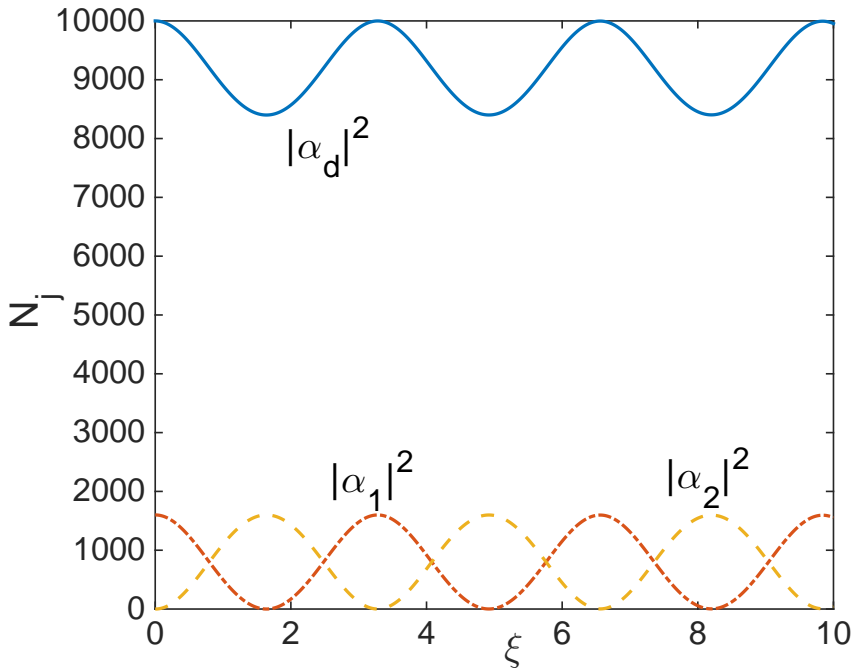}
\caption{(colour online) Field intensities for $|\alpha_{d}(0)|^{2}=10^{4}$, $|\alpha_{1}(0)|^{2}=1.6\times 10^{3}$, $|\alpha_{2}(0)|^{2}=0$, and $\kappa=10^{-2}$. The x-coordinate is a dimensionless interaction time, $\xi=\kappa |\alpha_{d}(0)|t$. All quantities plotted in this work are dimensionless.}
\label{fig:Ntrav}
\end{figure}

This population transfer is shown in Fig.~\ref{fig:Ntrav}, with results which were averaged over $1.5\times 10^{6}$ stochastic trajectories, with convergence to less than the line widths. In this case, we took the field at $\omega_{1}$ as being initially populated. If the field at $\omega_{2}$ is initially populated instead, the only change is that the lines for the upper and lower frequency fields swap places. In this simulation, we have taken the initial states as being either coherent or vacuum. While other quantum initial states are possible~\cite{WPstates}, the effect of these is outside the scope of this work.
The process can be thought of as analogous to Rabi cycling between the upper and lower frequency fields, mediated by the field at $\omega_{d}$. 

The quantum statistics of the fields are also of interest, with intermode correlations developing during the dynamics. To analyse these correlations, we consider the Reid Einstein-Podolsky-Rosen (EPR) steering correlations~\cite{EPRMDR}, which detect the presence of bipartitions exhibiting the EPR paradox~\cite{EPR}. These correlations use the product of two inferred quadrature variances,
\begin{eqnarray}
V_{inf}(\hat{X}_{ij}) &=& V(\hat{X}_{i})-\frac{\left[V(\hat{X}_{i},\hat{X}_{j})\right]^{2}}{V(\hat{X}_{j})}, \nonumber \\
V_{inf}(\hat{Y}_{ij}) &=& V(\hat{Y}_{i})-\frac{\left[V(\hat{Y}_{i},\hat{Y}_{j})\right]^{2}}{V(\hat{Y}_{j})},
\label{eq:VXYinf}
\end{eqnarray}
and the Heisenberg Uncertainty Principal to demonstrate the paradox whenever  $V_{inf}(\hat{X}_{ij})V_{inf}(\hat{Y}_{ij}) < 1$. In the modern language of steering introduced by Wiseman \etals~\cite{Wisesteer}, $V_{inf}(\hat{X}_{ij})V_{inf}(\hat{Y}_{ij}) < 1$ means that measurements on mode $j$ can steer the ensemble of possible measurements to be made on mode $i$. The variances and covariances follow the usual definitions, $V(A)=\langle A^{2}\rangle-\langle A\rangle^{2}$ and $V(A,B)=\langle AB\rangle-\langle A\rangle\langle B\rangle$.

We will use the notation EPR$_{jk}$ as the product of the $\hat{X}_{jk}$ and $\hat{Y}_{jk}$ inferred variances. The directionality of the paradox is recognised in the fact that EPR$_{jk}$, where mode $j$ is steered by measurements of mode $k$, is not always equal to EPR$_{kj}$. The situation where one of these is less than one while the other is more than one is known as Gaussian asymmetric steering~\cite{SFG,sapatona,Natexp,oneway,meu}. We note that our quadratures are defined as $\hat{X}_{j}=\hat{a}_{j}+\hat{a}_{j}^{\dag}$ and $\hat{Y}_{j}=-i(\hat{a}_{j}-\hat{a}_{j}^{\dag})$ and that other definitions are possible, but that these merely change the numerical value of the inequality. Because the EPR steerable states are a strict subset of the entangled states, both symmetric and asymmetric steering demonstrate that the two modes concerned are fully bipartite entangled.

\begin{figure}[tbhp]
\includegraphics[width=0.75\columnwidth]{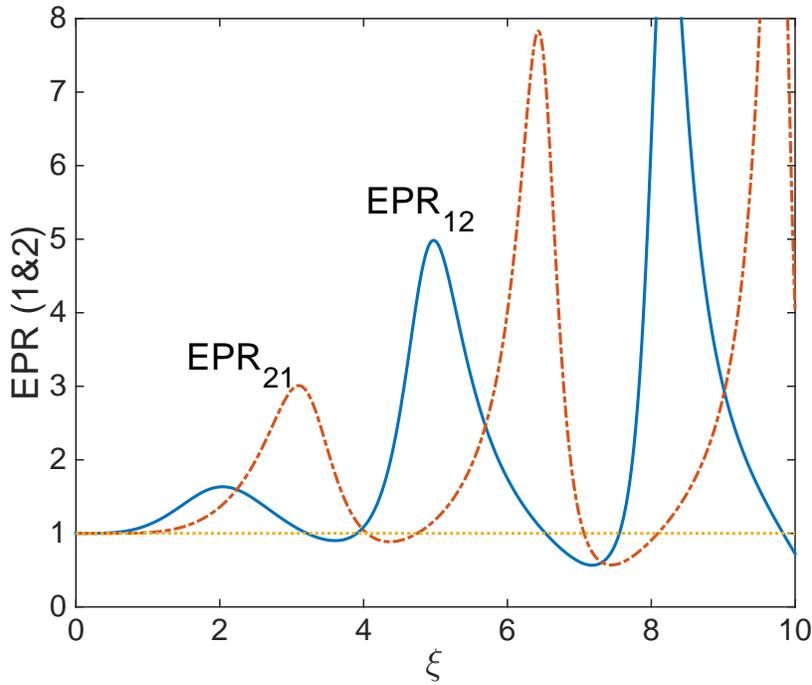}
\caption{(colour online) The Reid EPR correlations EPR$_{12}$ and EPR$_{21}$ for the same parameters and initial conditions as in Fig.~\ref{fig:Ntrav}. The dotted line at one is a guide to the eye.}
\label{fig:EPR12trav}
\end{figure}

\begin{figure}[tbhp]
\includegraphics[width=0.75\columnwidth]{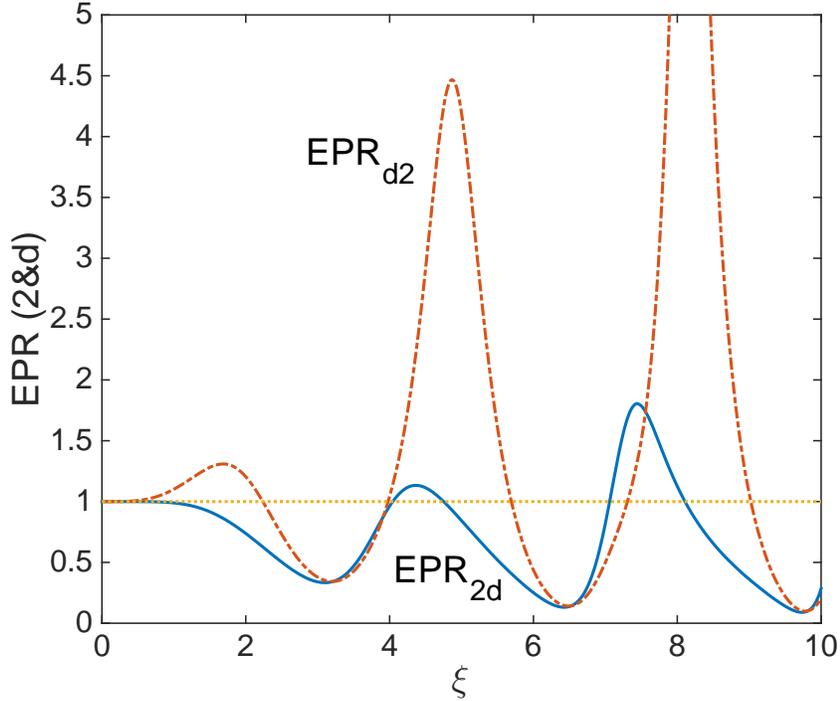}
\caption{(colour online) The Reid EPR correlations EPR$_{2d}$ and EPR$_{d2}$ for the same parameters and initial conditions as in Fig.~\ref{fig:Ntrav}. The dotted line at one is a guide to the eye.}
\label{fig:EPR2dtrav}
\end{figure}

The first of these we investigate is the possibility of steering between the low and high frequency modes, as shown in Fig.~\ref{fig:EPR12trav}. We fnd that the EPR correlations take some time before they initially fall below one, which is expected since they must develop through the interaction of the two modes. We see that the violation of the inequalities is cyclic for both partitions, with each minima being smaller than the previous one. There are interaction times where the steering is clearly aysmmetric and others where it is symmetric. Both the extent to which the inequalities are violated and the excess noise above the independent coherent state level of one increase with each cycle. We note here that these results are from a simplified model which neglects effects such as losses and dispersion in the nonlinear material, so that they become less likely to be experimentally reproducible as $\xi$ increases. However, they are a good indication of what is possible in terms of quantum correlations from this interaction Hamiltonian. When we consider the intracavity process in the next section, we fully expect the results to be more reproducible in the laboratory.

The low and high frequency modes also become entangled with the difference mode as the interaction proceeds. The EPR steering results for the high frequency field with the difference field are shown in Fig.~\ref{fig:EPR2dtrav}. The results for the bipartition of the difference field and the low frequency field are qualitatively similar.  Again we find regions of both symmetric and asymmetric steering. We will now proceed to examine the quantum correlations of the intracavity configuration.

\section{Intracavity Hamiltonian and Output Spectra}
\label{sec:cav}

When the interaction between the three fields and the nonlinear material takes place inside a pumped Fabry-Perot cavity, the effective description is in terms on two different partial Hamiltonians and a Liouvillian term. The first term is the interaction Hamiltonian of Eq.~\ref{eq:intham}. The cavity pumping is described by the Hamiltonian
\begin{equation}
{\cal H}_{pump} = i\hbar\left(\epsilon_{d}\hat{a}_{d}^{\dag}+\epsilon_{1}\hat{a}_{1}^{\dag}+\epsilon_{2}\hat{a}_{2}^{\dag} \right) + h.c,
\label{eq:Hpump}
\end{equation}
where the $\epsilon_{j}$ represent coherent input fields at frequency $\omega_{j}$. Note that we are considering that all fields are resonant with the cavity, although including any detunings is a simple procedure if required~\cite{detune}.

The damping of the cavity fields into a zero temperature Markovian reservoir is described by the Lindblad superoperator 
\begin{equation}
{\cal L}\rho = \sum_{i}\gamma_{i}\left(2\hat{a}_{i}\rho\hat{a}_{i}^{\dag}-\hat{a}_{i}^{\dag}\hat{a}_{i}\rho-\rho\hat{a}_{i}^{\dag}\hat{a}_{i} \right),
\label{eq:Lindblad}
\end{equation}
where $\rho$ is the system density matrix and $\gamma_{i}$ is the cavity loss rate at $\omega_{i}$.

In a doubled phase space, this again results in six coupled stochastic positive-P representation differential equations,
\begin{eqnarray}
\frac{d\alpha_{1}}{dt} &=&\epsilon_{1}-\gamma_{1}\alpha_{1}+ \kappa\alpha_{d}^{+}\alpha_{2}+\sqrt{\frac{\kappa\alpha_{2}}{2}}(\eta_{1}+i\eta_{3}), \nonumber \\
\frac{d\alpha_{1}^{+}}{dt} &=&\epsilon_{1}^{\ast}-\gamma_{1}\alpha_{1}^{+} +  \kappa\alpha_{d}\alpha_{2}^{+}+\sqrt{\frac{\kappa\alpha_{2}^{+}}{2}}(\eta_{2}+i\eta_{4}), \nonumber \\
\frac{d\alpha_{d}}{dt} &=& \epsilon_{d}-\gamma_{d}\alpha_{d}+ \kappa\alpha_{1}^{+}\alpha_{2}+\sqrt{\frac{\kappa\alpha_{2}}{2}}(\eta_{1}-i\eta_{3}), \nonumber \\
\frac{d\alpha_{d}^{+}}{dt} &=& \epsilon_{d}^{\ast}-\gamma_{d}\alpha_{d}^{+}+ \kappa\alpha_{1}\alpha_{2}^{+}+\sqrt{\frac{\kappa\alpha_{2}^{+}}{2}}(\eta_{2}-i\eta_{4}), \nonumber \\
\frac{d\alpha_{2}}{dt} &=& \epsilon_{2}-\gamma_{2}\alpha_{2} -\kappa\alpha_{1}\alpha_{d}, \nonumber \\
\frac{d\alpha_{2}^{+}}{dt} &=& \epsilon_{2}^{\ast}-\gamma_{2}\alpha_{2}^{+} -\kappa\alpha_{1}^{+}\alpha_{d}^{+},
\label{eq:PPcav}
\end{eqnarray}
where the $\alpha_{j}$ are the intracavity equivalents of the variables of Eq.~\ref{eq:Pplustravel}.

When nonlinear optical media are held inside a pumped optical cavity, the measured observables are usually the output spectral correlations, which are accessible using homodyne measurement techniques~\cite{mjc}. These are readily calculated in the steady-state by treating the system as an Ornstein-Uhlenbeck process~\cite{SMCrispin}. In order to do this, we begin by expanding the positive-P variables into their steady-state expectation values plus delta-correlated Gaussian fluctuation terms, e.g.
\begin{equation}
\alpha_{ss} \rightarrow \langle\hat{a}\rangle_{ss}+\delta\alpha.
\label{eq:fluctuate}
\end{equation}
Given that we can calculate the $\langle\hat{a}\rangle_{ss}$, we may then write the equations of motion for the fluctuation terms. The resulting equations are written for the vector of fluctuation terms as
\begin{equation}
\frac{d}{dt}\delta\vec{\alpha} = -A\delta\vec{\alpha}+Bd\vec{W},
\label{eq:OEeqn}
\end{equation}
where $A$ is the steady-state drift matrix, $B$ is found from the factorisation of the diffusion matrix of the original Fokker-Planck equation whcih was mapped onto our stochastic differential equations, $D=BB^{T}$, with the steady-state values substituted in, and $d\vec{W}$ is a vector of Wiener increments. As long as the matrix $A$ has no eigenvalues with negative real parts, this method may be used to calculate the intracavity spectra via
\begin{equation}
S(\omega) = (A+i\omega)^{-1}D(A^{\mbox{\small{T}}}-i\omega)^{-1},
\label{eq:Sout}
\end{equation}
from which the output spectra are calculated using the standard input-output relations~\cite{mjc}. For the results presented here, the real parts of the eigenvalues of $A$ were always positive, and stochastic integration showed that all three fields entered stationary steady states.

In this case the semi-classical equations are found by removing the noise terms from Eq.~\ref{eq:PPcav}, leaving three equations since $\alpha_{i}^{+}$ becomes $\alpha_{i}^{\ast}$ without the noise terms. Solving the resulting equations for the steady state with $\epsilon_{1}=\epsilon_{2}=0$ gives us $\alpha_{d}=\epsilon_{d}/\gamma_{d}$ and $|\alpha_{1}|^{2}=-|\alpha_{2}|^{2}$, which necessarily means $\alpha_{1}=\alpha_{2}=0$. This is again to be expected, since any other solutions would violate energy conservation. 

Setting $\epsilon_{1}\neq 0$ and $\epsilon_{2}=0$ (or the other way around), we find that the semi-classical equations for the steady-state are difficult to solve analytically in general, giving us fifth order polynomials. We will therefore proceed numerically in what follows.      
The drift matrix $A$ is found as 
\begin{equation}
A =
\begin{bmatrix}
\gamma_{d} & 0 & 0 & -\kappa\alpha_{2} & - \kappa\alpha_{1}^{\ast} & 0 \\
0 & \gamma_{d} & -\kappa\alpha_{2}^{\ast} & 0 & 0 & -\kappa\alpha_{1} \\
0 & -\kappa\alpha_{2} & \gamma_{1} & 0 & -\kappa\alpha_{d}^{\ast} & 0 \\
-\kappa\alpha_{2}^{\ast} & 0 & 0 & \gamma_{1} & 0 & -\kappa\alpha_{d} \\
\kappa\alpha_{1} & 0 & \kappa\alpha_{d} & 0 & \gamma_{2} & 0 \\
0 & \kappa\alpha_{1}^{\ast} & 0 & \kappa\alpha_{d}^{\ast} & 0 & \gamma_{2}
\end{bmatrix},
\label{eq:Amat}
\end{equation}
and $D$ is a $6\times 6 $ matrix with
\begin{eqnarray}
D(1,3) &=& D(3,1) = \kappa\alpha_{2}, \nonumber \\
D(2,4) &=& D(4,2) = \kappa\alpha_{2}^{\ast},
\label{eq:Dmat}
\end{eqnarray}
and all other elements being zero.
In the above two equations, the $\alpha_{j}$ should be read as the steady-state values.
Because we have parametrised our system using $\gamma_{d}=1$, the frequency $\omega$ is in units of $\gamma_{d}$. $S(\omega)$ then gives us products such as $\delta\alpha_{i}\delta\alpha_{j}$ and  $\delta\alpha_{i}^{\ast}\delta\alpha_{j}^{\ast}$, from which we construct the output variances and covariances for modes $i$ and $j$ as
\begin{equation}
S^{out}(X_{i},X_{j}) = \delta_{ij}+\sqrt{\gamma_{i}\gamma_{j}} \left(S_{ij}+S_{ji}\right).
\label{eq:Sout}
\end{equation}

\begin{figure}[tbhp]
\includegraphics[width=0.75\columnwidth]{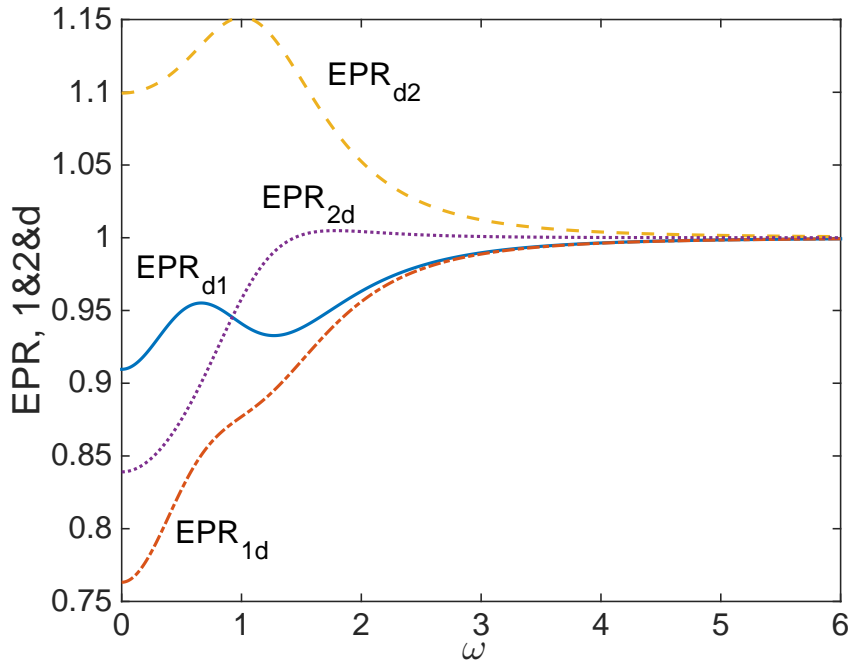}
\caption{(colour online) The output Reid EPR correlations between the field at $\omega_{d}$ and the other two, for $\epsilon_{d}=100$, $\epsilon_{1}=50$, $\kappa=10^{-2}$ and $\gamma_{d}=\gamma_{1}=\gamma_{2}=1$. The frequency axis is in units of $\gamma_{d}$.}
\label{fig:EPRSFGcav}
\end{figure}

The Reid EPR correlations, shown in Fig.~\ref{fig:EPRSFGcav} for $\epsilon_{d}=100$, $\epsilon_{1}=50$, $\kappa=10^{-2}$ and $\gamma_{d}=\gamma_{1}=\gamma_{2}=1$, show us that the difference mode is entangled with both modes $1$ and $2$, with the bipartition of $1$ and $d$ exhibiting symmetric steering while $2$ and $d$ exhibit asymmetric steering, where measurements on $d$ can steer mode $2$, but not vice-versa.  
For these parameters, we find no EPR steering between modes $1$ and $2$. The Duan-Simon method~\cite{Duan,Simon}, based on the positivity of the partially transposed density matrix for a physical system and commonly used to detect bipartite entanglement, also fails to indicate that these two modes are either inseparable or entangled. We note that all three fields are macroscopically occupied in the steady-state, with $|\alpha_{d}|^{2}=8,712$,  $|\alpha_{1}|^{2}=714$, and $|\alpha_{2}|^{2}=622$.

\begin{figure}[tbhp]
\includegraphics[width=0.75\columnwidth]{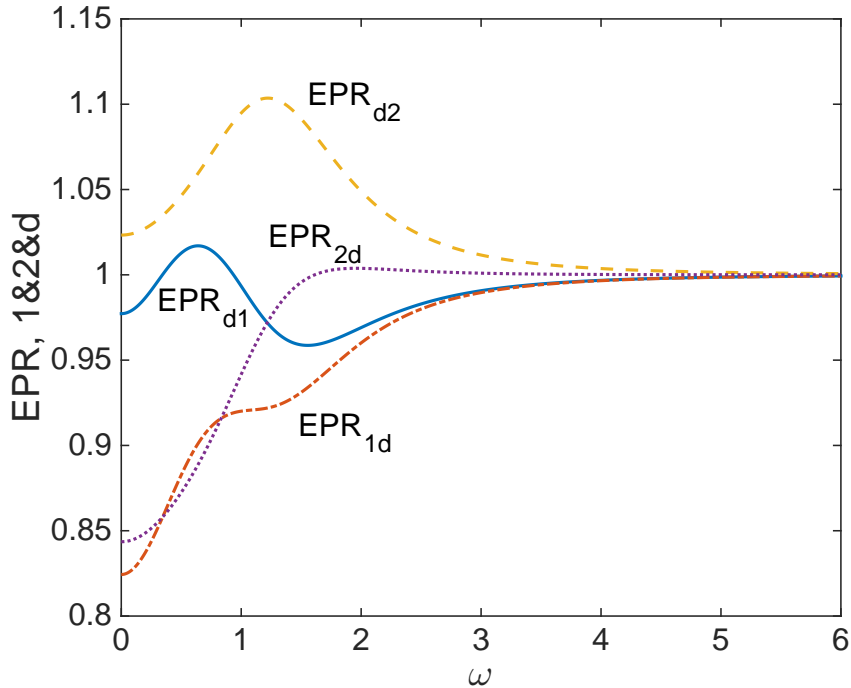}
\caption{(colour online) The output Reid EPR correlations between the field at $\omega_{d}$ and the other two, for $\epsilon_{d}=100$, $\epsilon_{2}=50$, $\kappa=10^{-2}$ and $\gamma_{d}=\gamma_{1}=\gamma_{2}=1$. The frequency axis is in units of $\gamma_{d}$.}
\label{fig:EPRNOPOcav}
\end{figure}

When the high frequency mode and the difference mode are both pumped, the results are qualitatively similar, as shown in Fig.~\ref{fig:EPRNOPOcav}. The same two modes exhibit asymmetric steering and again we find no sign of entanglement between the high and low frequency fields, using either the Reid EPR or Duan-Simon measures. In this configuration, we find the steady state intensities  $|\alpha_{d}|^{2}=12,100$,  $|\alpha_{1}|^{2}=623$, and $|\alpha_{2}|^{2}=556$. The difference mode has a higher steady-state intensity in this case because more energy is entering the cavity due to the pumping at the higher frequency, so more is available to populate it above the non-interacting steady-state value of $|\alpha_{d}|^{2}=|\epsilon|^{2}/\gamma_{d}^{2}$.  

\section{Conclusions}
\label{sec:conclusions}

We have shown that a system which was previously analysed classically and claimed to have no quantum counterpart can in fact be given a full quantum description that does not violate energy conservation. The process will proceed as long as either the high frequency field or the two lower frequency ones either have some initial population (in the travelling wave case), or are pumped (in the intracavity case). Using the appropriate quantum Hamiltonian, we have shown that the proposed mechanism of difference parametric amplification can be thought of as a number of three-wave mixing processes, all of which are well known. 

We have analysed the mean-field dynamics and the quantum correlations needed to prove bipartite entanglement between the modes, in both the travelling wave and intracavity cases. For the parameters considered, we found all three bipartitions to be entangled in the travelling wave case, exhibiting asymmetric steering at some interaction times. In the intracavity case, the upper and lower frequency modes were found to be entangled with the difference mode, but not with each other.

\end{document}